# Superconducting ternary compounds Li-X-B (X=Mo, W) within the mild pressure range: First-principles predictions

Bangshuai Zhu[1#], Juefei Wu[1#*], Dexi Shao[2], Junjie Wang[4, 5], Yu Han[4, 5], Cuiying Pei[1], Qi Wang[1,3], Jian Sun[4,5], Yanpeng Qi[1,3,6*].

1. State Key Laboratory of Quantum Functional Materials, School of Physical Science and Technology, ShanghaiTech University, Shanghai 201210, China

2. School of Physics, Hangzhou Normal University, Hangzhou 311115, China

3. ShanghaiTech Laboratory for Topological Physics, ShanghaiTech University, Shanghai 201210, China

4. National Laboratory of Solid State Microstructures, School of Physics and Collaborative Innovation Center of Advanced Microstructures, Nanjing University, Nanjing 210093, China

5. Collaborative Innovation Center of Advanced Microstructures, Nanjing 210093, China

6. Shanghai Key Laboratory of High-resolution Electron Microscopy, ShanghaiTech University, Shanghai 201210, China

[#]These authors contributed equally.

[*]Correspondence should be addressed to J.F.W. (wujf@shanghaitech.edu.cn) or Y.P.Q. (qiyp@shanghaitech.edu.cn)

## Abstract

Among the superconducting hydrides under high pressure, a number of studies concentrate on the ternary compounds to explore unique superconductors, which are capable of reducing the stable pressure and maintain superconductivity. In this work, to verify our proposed strategy of ternary composition lines (TCLs) to explore ternary compounds, we combined the first-principles calculations and crystal structure predictions to study the ternary compounds Li-X-B (X=Mo, W) under high pressure. After calculations along five and four TCLs in Li-W-B and Li-Mo-B, respectively, five Li-W-B compounds and four Li-Mo-B compounds were predicted. The compositions of $LiWB_4$, $Li_4MoB_2$ and $LiMo_2B_2$ could be thermodynamically stable under high pressure, and $Li_2WB_6$ is around 0.02 eV/atom above the convex hull at 0 GPa, which has potential for synthesizing. Both of the predicted $Li_2WB_6$ *P*6/*mmm* and $Li_2WB_4$ *R*-3*m* are superconducting and their $T_c$ are around 11 K, which are similar to the $T_c$ of $WB_2$ *P*6/*mmm* around 100 GPa. An anomalous increase of $T_c$ was found in $Li_4MoB_2$ *C*2/*m* upon compression. We carried out full ternary search (FTS) to evaluate the validity of the TCLs strategy in Li-W-B system at 0 GPa. Our results are helpful for understanding the phase diagram of Li-X-B (X=Mo, W) under high pressure and the introducing of Li atoms provide candidate structures to reduce the measured stable pressure from ~100 GPa in $WB_2$ *P*6/*mmm* to 0 GPa. Meanwhile, we preliminary validate the strategy of TCLs in structure predictions and we expect to improve this strategy in the future, shedding light on the studies of ternary compounds.

## I. Introduction

The successful predictions and synthesizing of $H_3S$ under high pressure [1-4] motivated the studies of high temperature superconductors in the compressed binary hydrides, such as $LaH_{10}$ [5, 6], $YH_9$ [7], $CaH_6$ [8-10], etc [11-15]. Despite the noticeable superconducting critical temperature ($T_c$) records, most of them are synthesized over 150 GPa, which is still challenging in the laboratory and confines the practical applications. Hence recent studies focus on the ternary hydrides, which have two orders-of-magnitude more structures than the binary compounds. This provides possibility to reduce the synthesizing pressure and retain the superconducting properties. In terms of experiments, the synthesizing pressure ($P_s$) of La-Y-H is 170-196 GPa with the measured $T_c$ ~253 K [16], while the $P_s$ ~113 GPa for (La, Ce)$H_{9-10}$ with the measured $T_c$ ~176 K [17, 18]. Examples combining synthesis and predictions include $LaBeH_8$ ($P_s$ 110-113 GPa, $T_c$ ~110 K) [19, 20] and (Y, Ca)$H_6$ ($P_s$ ~170 GPa, $T_c$ ~224 K) [21, 22]. Besides, more ternary hydrides are predicted to be dynamically stable within the mild pressure range ($P \leq 50$ GPa) [23], such as $KB_2H_8$ (12 GPa, $T_c$ ~134 K) [24], $ThBeH_8$ (7 GPa, $T_c$ ~113 K) [25], $ThSiH_7$ ($T_c$ ~134 K) [26] etc [27-29]. The typical ternary hydrides could reduce the synthesizing pressure to around 100 GPa and the predicted structures are candidate structures for further pressure reduction.

In addition to hydrides, light element compounds based on Li, B, P etc also have advantages for superconductivity such as high Debye temperature and strong electron-phonon coupling (EPC), and Ref. 30 suggests that these compounds are promising to seek breakthroughs. For instance, the synthesized borides like $\alpha$-$MoB_2$ ($T_c$ ~32 K) [31, 32], $WB_2$ ($T_c$ ~15-17 K) [33, 34], $MgB_2$ ($T_c$ ~39 K) [35], the predicted borides contain $CaB_2$ ($T_c$ ~48 K) [36], $KB_7$ ($T_c$ ~26 K) [37], $CeB_2$ ($T_c$ ~29 K) and $CeB_8$ ($T_c$ ~27 K) [38], etc [39-42]. The predicted sulfides include $Sn_3S_4$ ($T_c$ ~22 K) [43], GeS ($T_c$ ~28 K) [44], $LuS_6$ ($T_c$ ~26 K) [45], etc [46-49]. The stable pressures for most of the light element compounds are below 100 GPa, and they can act as the precursors for exploring ternary compounds with unexpected properties [50], such as the prediction of Li-Mn-B system that $LiMnB_4$ has high hardness of ~39 GPa and $Li_2MnB_8$ has $T_c$ ~15 K at 50 GPa [51]. Considering the distinctive role of 4*d* and 5*d* electrons in typical hydrides ($LaH_{10}$ and $YH_9$) [5-7] and borides ($\alpha$-$MoB_2$ and $WB_2$) [31-34] along with the similarity between Li and H atoms, it could be intriguing to explore unique ternary compounds combing Li and X-B (X=Mo, W) systems under high pressure.

In terms of theoretical predictions, there are in general two common strategies to calculate the ternary convex hull under high pressure. One strategy utilizes the random crystal structure searching methods or high-throughput screening supported by crystal structures database [25, 50, 52-56], which scan the whole ternary convex hull but can be computationally expensive and time-consuming. The other strategy concentrates on certain characteristics, such as the specific compositions, the typical prototype structures and the elements substitution [16, 19, 21, 23, 24, 26-29, 51], which efficiently evaluate structures with unique properties but the lack of stoichiometries could lead the deficiency in the thermodynamic stability. Thus, we propose a strategy termed as ternary composition lines (TCLs), which constructs specific paths and focuses on the polyhedron of the ternary convex hull. The TCLs are based on two key assumptions. The first is that a predicted ternary compound could be on the specific line of the ternary convex hull, considering the possible synthesizing path. The second is that current ternary convex hull reflects the potential distribution of the stable and meta-stable ternary compositions. Based on these two assumptions, we could construct TCLs and predict stable or meta-stable ternary compounds, which is helpful to preliminary locate the areas of the energy minimums in the ternary convex hull with limited computational resources.

In this work, we preliminary verified the strategy of TCLs by combining the first-principles calculations and crystal structure predictions to investigate the ternary compounds in Li-X-B (X=Mo, W) systems within 60 GPa. We tested five TCLs in the Li-W-B system and four TCLs in the Li-Mo-B system, and predicted five unique structures in Li-W-B and four structures in Li-Mo-B. The compositions of $LiWB_4$, $Li_4MoB_2$ and $LiMo_2B_2$ could be stable within corresponding pressure ranges. In particular, the predicted $Li_2WB_6$ *P*6/*mmm* and $Li_2WB_4$ *R*-3*m* are meta-stable from 0 GPa and 20 GPa, respectively, and the EPC calculations indicate similar superconductivity mechanism as $WB_2$ *P*6/*mmm*. The $T_c$ values of $Li_2WB_6$ *P*6/*mmm* and $Li_2WB_4$ *R*-3*m* are about 11 K. The enthalpy difference relative to the convex hull at 0 GPa for $Li_2WB_6$ *P*6/*mmm* is less than 0.02 eV/atom, which has higher potential for synthesizing. Then, we performed a full ternary search (FTS) of Li-W-B at 0 GPa to evaluate the validity of the strategy of TCLs. Besides, we also find an abnormal increase of $T_c$ in $Li_4MoB_2$ *C*2/*m* from 0 GPa to 40 GPa.

## II. Methods

### A. The strategy of ternary composition lines (TCLs).

According to Ref. 57, most of the compounds adopt the compositions of $(AB_k)_x(AC_l)_y$ or $A_x(BC_m)_y$ ($AB_k$, $AC_l$ and $BC_m$ are stable binary compounds, $A_x$ represents pure elements) based on the octet rule. Hence one could accelerate the exploration of the chemical space by combining the pseudobinary joints and the algorithm like co-evolution [58]. Moreover, from the experimental perspective, a predicted ternary compound has higher potential for synthesizing if it could be on the certain line of the ternary convex hull, such as the superconducting hydrides under high pressure [17-22, 52]. This is the first assumption to construct the TCLs and we can enrich the compositions by considering the meta-stable compounds matching the threshold of 0.05 eV/atom [59].

As for the construction of TCLs, we suggest to concentrate on the polyhedron of the ternary convex hull. In the binary system of X-Y (X and Y are pure elements), the composition distribution has respective preference under high pressure, such as X or Y rich compounds or certain stoichiometry [5, 8, 37, 38, 40, 41, 43-48, 60, 61]. By analogy, we propose the second assumption that the ternary convex hull constructed by known compounds could partly reflect the potential areas containing stable and meta-stable ternary compositions. This is supported by predictions of ternary systems like W-Cr-B, Hf-Ta-C [57] and La-Sc-H [52] under high pressure, where the composition distribution has close relationships with the pseudobinary joints involving stable binary compounds.

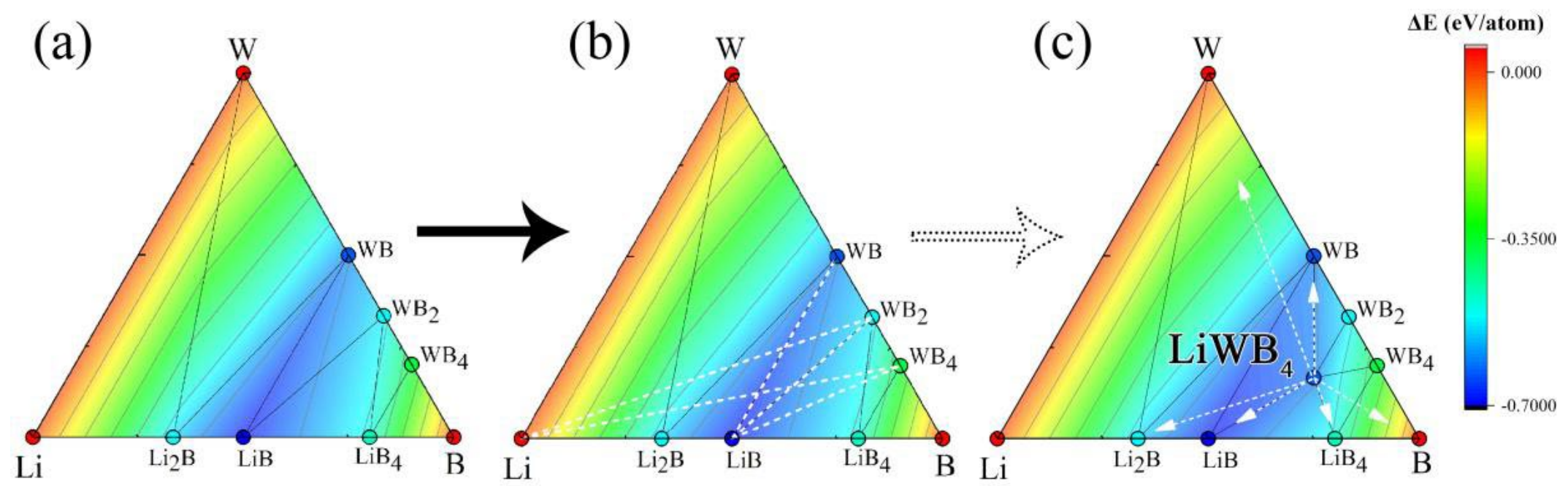

FIG. 1 The ternary composition line (TCL) in the ternary convex hulls of Li-W-B at 40 GPa. (a) The initial ternary convex hull. (b) The 1st generation of TCLs. (c) The predicted ternary composition with thermodynamic stability and potential 2nd generation of TCLs.

Taking the Li-W-B system for illustration, we construct the initial convex hull at 40 GPa [Fig. 1(a)] based on the reported binary compounds and simple substance [62-69], from which we can find the low energy region distributing along the line of LiB-WB. Based on our two assumptions, we suggest to prioritize pseudobinary joints across or along this low energy region and choose five lines including Li-$WB_2$, Li-$WB_4$, LiB-WB, LiB-$WB_2$ and LiB-$WB_4$, as plotted in Fig. 1 (b). We name these joints based on the polyhedron of the initial ternary convex hull as the 1st generation of TCLs. Meanwhile, four 1st generation of TCLs are picked out in Li-Mo-B system including Li-MoB, Li-$MoB_2$, LiB-MoB and LiB-$MoB_2$ [Fig. S1] [70]. Afterwards, we predicted a stable composition of $LiWB_4$ at 40 GPa, based on which the potential 2nd generation of TCLs could be constructed in Fig. 1 (c), and higher generations of TCLs could follow analogous routine.

**B. Structure searches and first-principles calculations.**

We performed the crystal structure searches combining the machine learning and graph theory assisted universal structure searcher (MAGUS) [71, 72] with Vienna *Ab-initio* Simulation Package (VASP) based on the density functional theory [73, 74] at 20 GPa and 50 GPa, respectively. The variable composition searches ascertained the compositions on each TCLs, which were further rechecked by the fixed composition searches. We utilized the VASP to calculate the enthalpy and electronic structures of the predicted phases. The exchange-correlation functional was the generalized gradient approximation (GGA) of Perdew, Burkey, and Ernzerhof formula [75] and the valence electrons were treated by the projector-augmented wave (PAW) approach [76]. The phonon spectra of the predicted structures were calculated by the finite displacement method implemented in the PHONOPY program package [77], which were rechecked by the QUANTUM-ESPRESSO (QE) package [78] using density-functional perturbation theory [79]. The EPC coefficients were calculated by QE and we used the Allen-Dynes modified McMillan formula [80] to estimate the $T_c$ values. More detailed calculation parameters can be seen in the Supplementary material [70].

## III. Results and Discussion

**A. Predictions of Li-W-B system.**

Following the structure searches along the five TCLs of Li-$WB_2$, Li-$WB_4$, LiB-WB, LiB-$WB_2$ and LiB-$WB_4$, we calculated the enthalpy and the phonon spectra of the predicted structures [Fig. S2, S3] [70], which construct the ternary convex hulls from 0 GPa to 60 GPa with the interval of 20 GPa, as shown in Fig. 2. The simple substance under corresponding pressures served as the references of the ternary convex hull [62, 63, 66-69]. We also conducted structure searches in Li-W system but the predicted structures host higher formation energy than Li plus W.

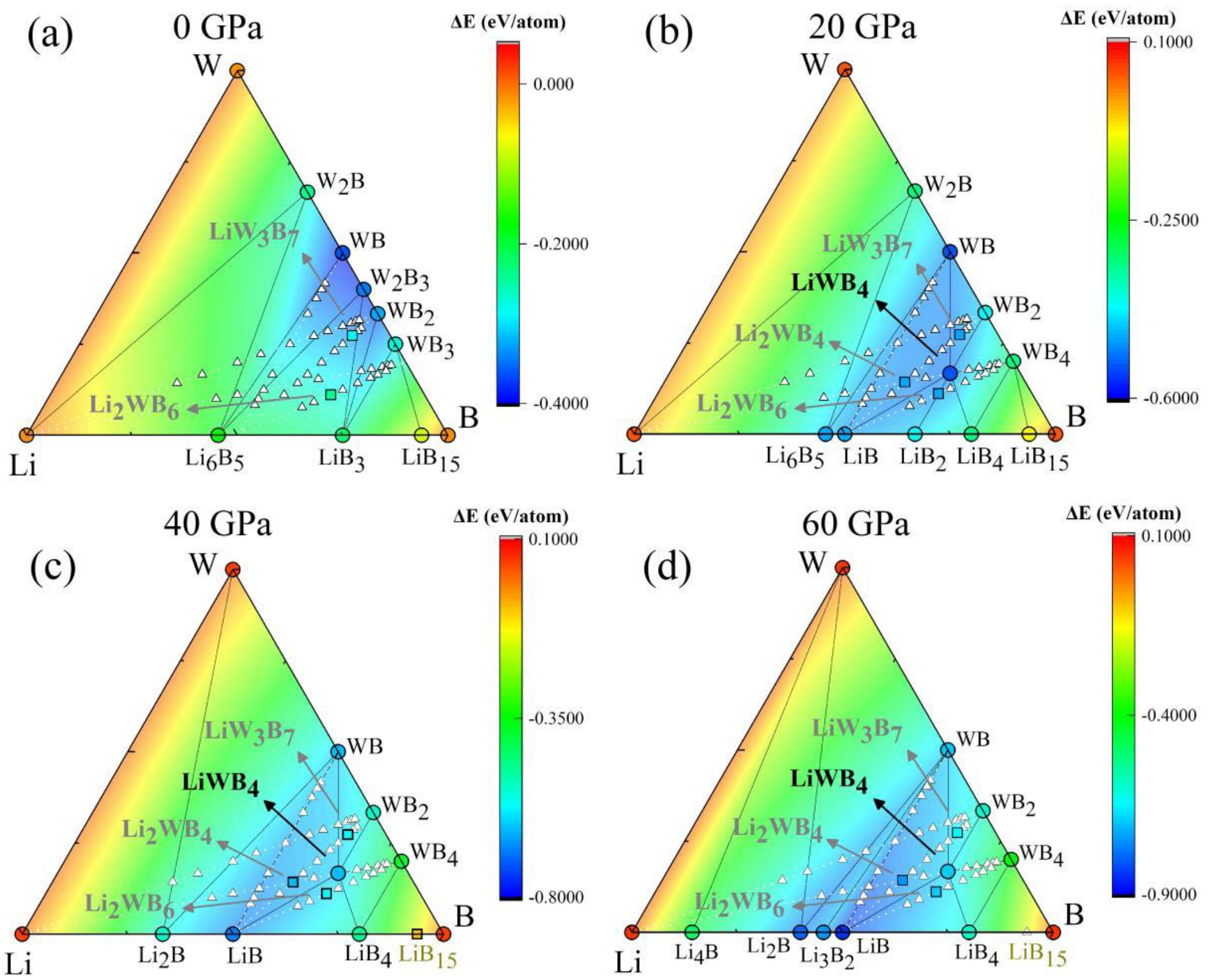

FIG. 2 (a)-(d) The ternary convex hulls of Li-W-B system under high pressures. The circle and black font are thermodynamically stable compositions, the square and grep font are thermodynamically meta-stable compositions.

We predicted four unique ternary compositions in the Li-W-B system. Among these compositions, $Li_2WB_6$ and $LiW_3B_7$ are meta-stable from 0 GPa [Fig. 2 (a)-(d)] and we predicted a pressure induced structural transition in $Li_2WB_6$. In particular, their enthalpy differences relative to the ternary convex hull are around 0.02 eV/atom at 0 GPa [Table. 1], which are comparable to the enthalpy difference in W-B compounds [50]. This indicates the potential for synthesis. Besides, the relative enthalpy of $Li_2WB_6$ is 0.04 eV/atom lower than LiB plus $WB_4$ at 0 GPa, suggesting the starting materials and potential pressure-temperature (P-T) condition for synthesis. Considering the dynamic stability by phonon spectra [Fig. S2] [70], the predicted composition of $Li_2WB_4$ is meta-stable after 20 GPa [Fig. 2 (b)-(d)], and the relative enthalpy difference decreases with pressure [Table. 1], which is less than 0.01 eV/atom at 60 GPa, implying that $Li_2WB_4$ lies on the ternary convex hull under higher pressure. Combining the calculations on enthalpy and phonon spectra [Fig. S3] [70], the predicted composition of $LiWB_4$ hosts both thermodynamic stability and dynamic stability after 10 GPa [Fig. 2 (b)-(d)], illustrating the potential for synthesizing under high pressure. The lattice parameters of the predicted Li-W-B phases are listed in Table. S1 [70].

TABLE. 1 The enthalpy difference of the predicted Li-W-B compounds relative to the ternary convex hull under high pressure.

| ΔEnthalpy (eV/atom) | | | | | |
|---|---|---|---|---|---|
| Phase | Space Group | Pressure (GPa) | | | |
| | | 0 | 20 | 40 | 60 |
| $Li_2WB_6$ | *P*6/*mmm* | 0.019 | 0.040 | - | - |
| | *C*2/*m* | - | - | 0.048 | 0.014 |
| $Li_2WB_4$ | *R*-3*m* | - | 0.046 | 0.027 | 0.006 |
| $LiW_3B_7$ | *Pmma* | 0.021 | 0.026 | 0.026 | 0.035 |
| $LiWB_4$ | *Cmcm* | - | 0 | 0 | 0 |

In terms of $Li_2WB_6$, the crystal structures of the low pressure phase $Li_2WB_6$ *P*6/*mmm* and the high pressure phase $Li_2WB_6$ *C*2/*m* are plotted in Fig. 3 (a) and (b). The $Li_2WB_6$ *P*6/*mmm* consists of the flat honeycomb B layers and Li-W layers. The Li and W atoms are above the center of the honeycomb B layers [Fig. 3(a)]. The $Li_2WB_6$ *C*2/*m* is also stacked by B layers and Li-W layers, while the sub lattices of Li and W atoms differ from that in $Li_2WB_6$ *P*6/*mmm*. As depicted in Fig. 3(b), we marked a Li-W layer sandwiched by B atoms. The configuration of Li and W atoms in the Li-W layer breaks the six-fold rotation symmetry in $Li_2WB_6$ *P*6/*mmm*, and we can observe a sliding between the neighboring Li-W layers in $Li_2WB_6$ *C*2/*m*. These features result in the buckling of the honeycomb boron layers [Fig. 3(b)]. In Fig. 3(c), we compared the enthalpy of the predicted two structures under high pressure, and the enthalpy value of $Li_2WB_6$ *C*2/*m* is lower than that of $Li_2WB_6$ *P*6/*mmm* beyond about 37 GPa, suggesting that $Li_2WB_6$ *C*2/*m* is more thermodynamically stable under higher pressure.

Besides, we calculated the electron localization function (ELF) [Fig. 3 (a) and (b)], band structures and partial density of states (PDOS) [Fig. 4 (a) and (b)] for both $Li_2WB_6$ *P*6/*mmm* and $Li_2WB_6$ *C*2/*m*. In $Li_2WB_6$ *P*6/*mmm*, the bonds between Li and B atoms are ionic, while the W atoms form covalent bonds with the B atoms [Fig. 3(a)]. The ELF suggests connection channels and isolation channels around W and Li atoms along the inter-layer direction, respectively [Fig. 3(a)]. In comparison, the connection areas surround the ionic bonds formed by Li atoms in $Li_2WB_6$ *C*2/*m* [Fig. 3(b)]. Besides, the ELF distribution of B layers reflect the honeycomb characters as well as the strong covalent B-B bonds in both $Li_2WB_6$ *P*6/*mmm* and $Li_2WB_6$ *C*2/*m* [Fig. 3 (a) and (b)]. The electronic structures indicate that the B-*p* electrons play a main role around the Fermi energy in $Li_2WB_6$ *P*6/*mmm*, while B-*p* and W-*d* electrons contribute equally around the Fermi energy in $Li_2WB_6$ *C*2/*m* [Fig. 4 (a) and (b)].

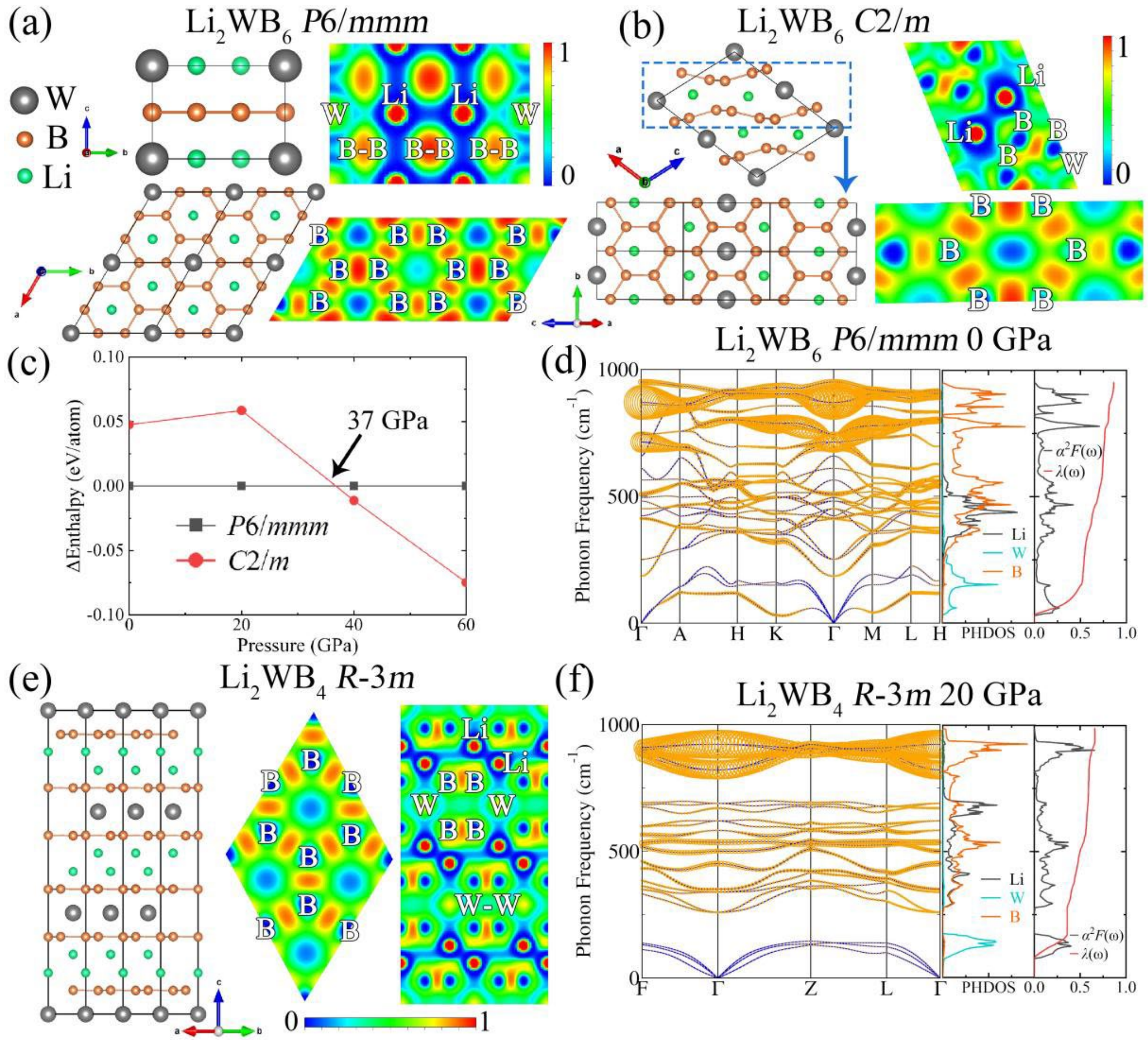

FIG. 3 (a) The crystal structure of the predicted phase $Li_2WB_6$ *P*6/*mmm* and ELF at (2-10) plane and the boron layer. (b) The crystal structure of the predicted phase $Li_2WB_6$ *C*2/*m* and ELF at (010) plane and the boron layer. (c) The enthalpy difference relative to $Li_2WB_6$ *P*6/*mmm* under high pressure. (d) The calculated phonon curves, PHDOS, Eliashberg spectral function $\alpha^2F(\omega)$, the electron phonon integral $\lambda(\omega)$ of $Li_2WB_6$ *P*6/*mmm* at 0 GPa. (e) The crystal structure and ELF at the boron layer and (-1-10) plane of the predicted phase $Li_2WB_4$ *R*-3*m*. (f) The calculated EPC properties of $Li_2WB_4$ *R*-3*m* at 20 GPa. The orange solid dots represent the EPC constant $\lambda$ and the radii are proportional to the strength.

Moreover, we calculated the EPC constants, projected phonon density of states (PHDOS), Eliashberg spectral function $\alpha^2F(\omega)$ and the electron-phonon integral $\lambda(\omega)$ to study the superconducting properties for $Li_2WB_6$ *P*6/*mmm* and $Li_2WB_6$ *C*2/*m* at 0 GPa and 40 GPa, respectively, while $Li_2WB_6$ *C*2/*m* is not superconducting. As shown in Fig. 3(d), the orange solid dots represent the EPC strength of the vibration modes. Combining the PHDOS, the W-dominant phonon modes occupy the acoustic branches within the frequency region below 250 cm$^{-1}$, and one of the acoustic branches is decorated with more orange dots, suggesting the contribution to the EPC. Meanwhile, the W atoms contribute about 54% of the integral $\lambda(\omega)$. These results suggest that the vibrations of W atoms are active in the EPC of $Li_2WB_6$ *P*6/*mmm*, which is similar to $WB_2$ *P*6/*mmm* under high pressure [81]. As plotted in Table. 2, the calculated EPC constant $\lambda$ is 0.96 with the estimated $T_c$ of 11.22 K for $Li_2WB_6$ *P*6/*mmm* at ambient pressure. This $T_c$ value is comparable to the experimental $T_c$ of 15 K in $WB_2$ under 100 GPa [33, 34], while the stable pressure is reduced to 0 GPa.

TABLE. 2 The calculated superconducting properties of the predicted $Li_2WB_6$ *P*6/*mmm*, $Li_2WB_4$ *R*-3*m* and $LiWB_4$ *Cmcm* under high pressure.

| Phase | Space Group | Pressure (GPa) | $\omega_{log}$ | $N_{Ef}$ | λ | $T_c$ (K) $\mu^* = 0.1$ |
|---|---|---|---|---|---|---|
| $Li_2WB_6$ | *P*6/*mmm* | 0 | 171.72 | 11.93 | 0.96 | 11.22 |
| $Li_2WB_4$ | *R*-3*m* | 20 | 292.24 | 11.02 | 0.74 | 11.44 |
| $LiWB_4$ | *Cmcm* | 10 | 580.59 | 14.96 | 0.35 | 0.96 |

As for the composition of $Li_2WB_4$, the predicted phase $Li_2WB_4$ *R*-3*m* [Fig. 3(e)] is analogous to the 3R transition metal dichalcogenides with the layer stacking of ABC. Differing from $Li_2WB_6$ *P*6/*mmm*, the Li atoms are independent of W atoms and separate the B-W-B layers in $Li_2WB_4$ *R*-3*m*. The ELF in the (-1-10) plane illustrates that Li atoms almost isolate the neighboring B-W-B layers and there is charge distribution within the W atoms layer. Besides, the PDOS indicate that W-*d* electrons dominate the electronic states around the Fermi energy [Fig. 4(c)]. These results imply that the intra-layer coupling between the B-W-B layers of $Li_2WB_4$ *R*-3*m* is relatively stronger than that of $Li_2WB_6$ *P*6/*mmm*, which could influence the superconducting properties. Thus, we calculated the EPC properties of $Li_2WB_4$ *R*-3*m* at 20 GPa, as displayed in Fig. 3(f). Despite W-dominant phonon modes below 180 $cm^{-1}$ contribute around 47% to the integral $\lambda(\omega)$, the vibrations of lighter atoms, particularly the B atoms with frequencies over 800 $cm^{-1}$, also play unambiguous role in the EPC,. This suggests that the coupling between the electrons of W atoms and the vibrations of B atoms is key to the EPC in $Li_2WB_4$ *R*-3*m*. The EPC constant $\lambda$ is 0.74 and the calculated $T_c$ is 11.44 K for $Li_2WB_4$ *R*-3*m* at 20 GPa [Table. 2]. The predictions of $Li_2WB_6$ *P*6/*mmm* and $Li_2WB_4$ *R*-3*m* suggest that the introduction of Li atoms in W-B compounds is capable of preserving superconducting properties and reducing the stable pressure.

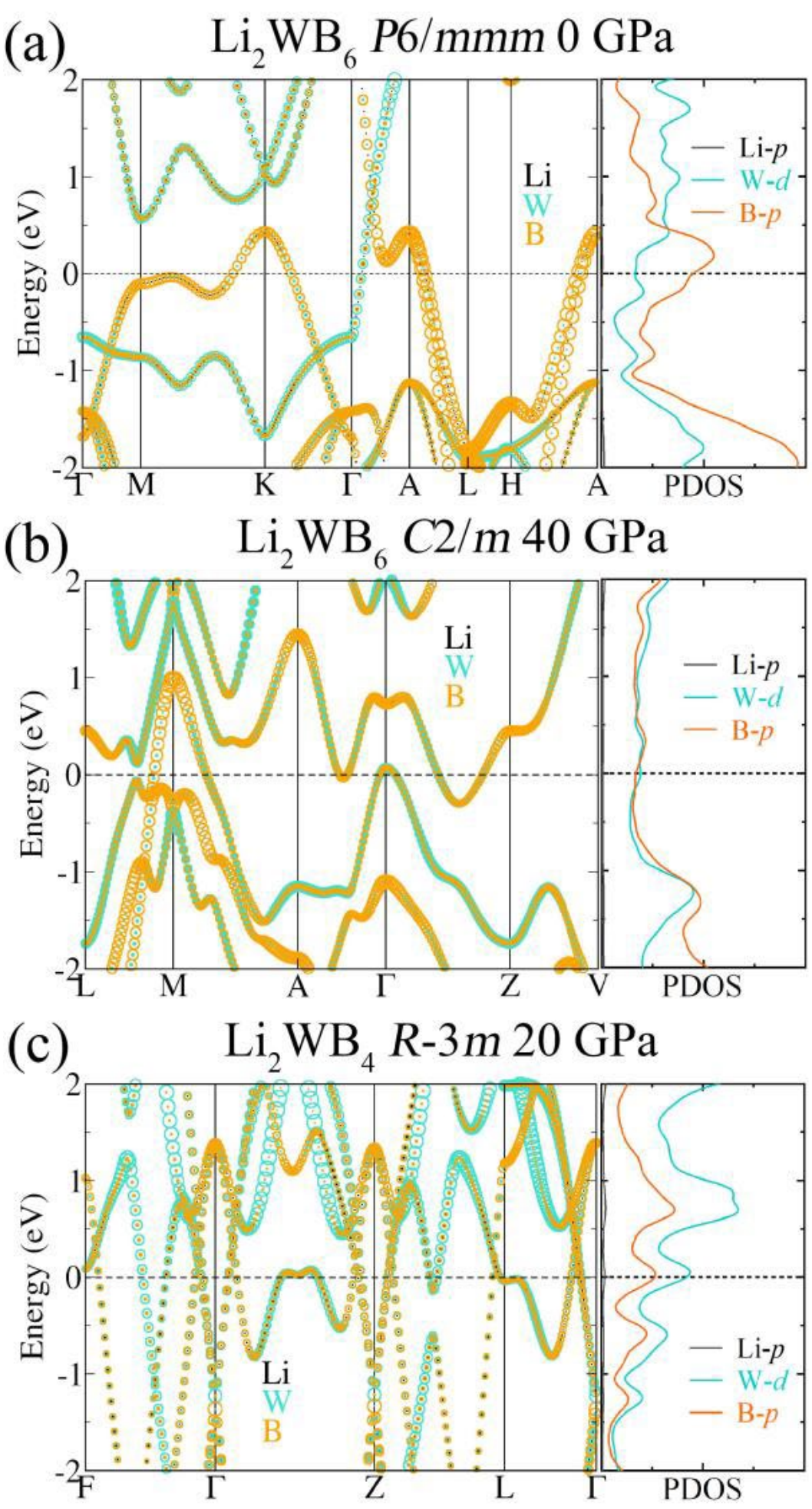

FIG. 4 The band structures and partial density of states (PDOS) of the predicted phases of (a) $Li_2WB_6$ *P*6/*mmm*, (b) $Li_2WB_6$ *C*2/*m* and (c) $Li_2WB_4$ *R*-3*m* under high pressure.

The predicted phase $LiWB_4$ *Cmcm* [Fig. 5(a)] is both thermodynamically and dynamically stable above 10 GPa [Fig. S3] [70]. The structure is stacked by the honeycomb B layers and Li-W layers. Although the crystal structure resembles that of the predicted $Li_2WB_6$ *P*6/*mmm*, the Li and W atoms follow the interval arrangement of Li-W-Li-W along the a-axis and b-axis [Fig. 4(a)]. This is in line with the ELF at the (010) plane that the isolated Li atoms block the inter-layer channel between the W atoms [Fig. 4(a)]. The calculations on the PDOS suggest that the W-*d* and B-*p* electrons have almost identical contribution around the Fermi energy [Fig. 5(d)], which is close to that of $Li_2WB_6$ *C*2/*m* [Fig. 4(b)]. The EPC calculations [Fig. 5(b)] suggest that we can divide two regions by 200 $cm^{-1}$ and part of the W-dominant vibrations in the low frequency region have certain contribution in the EPC. The contribution of W atoms to the integral $\lambda(\omega)$ is about 30%, which is lower than that of $Li_2WB_6$ *P*6/*mmm* and $Li_2WB_4$ *R*-3*m*. Besides, $LiWB_4$ *Cmcm* exhibits relatively weak superconductivity with the $\lambda$ value of 0.35 and the predicted $T_c$ value of 0.96 K at 10 GPa [Table. 2].

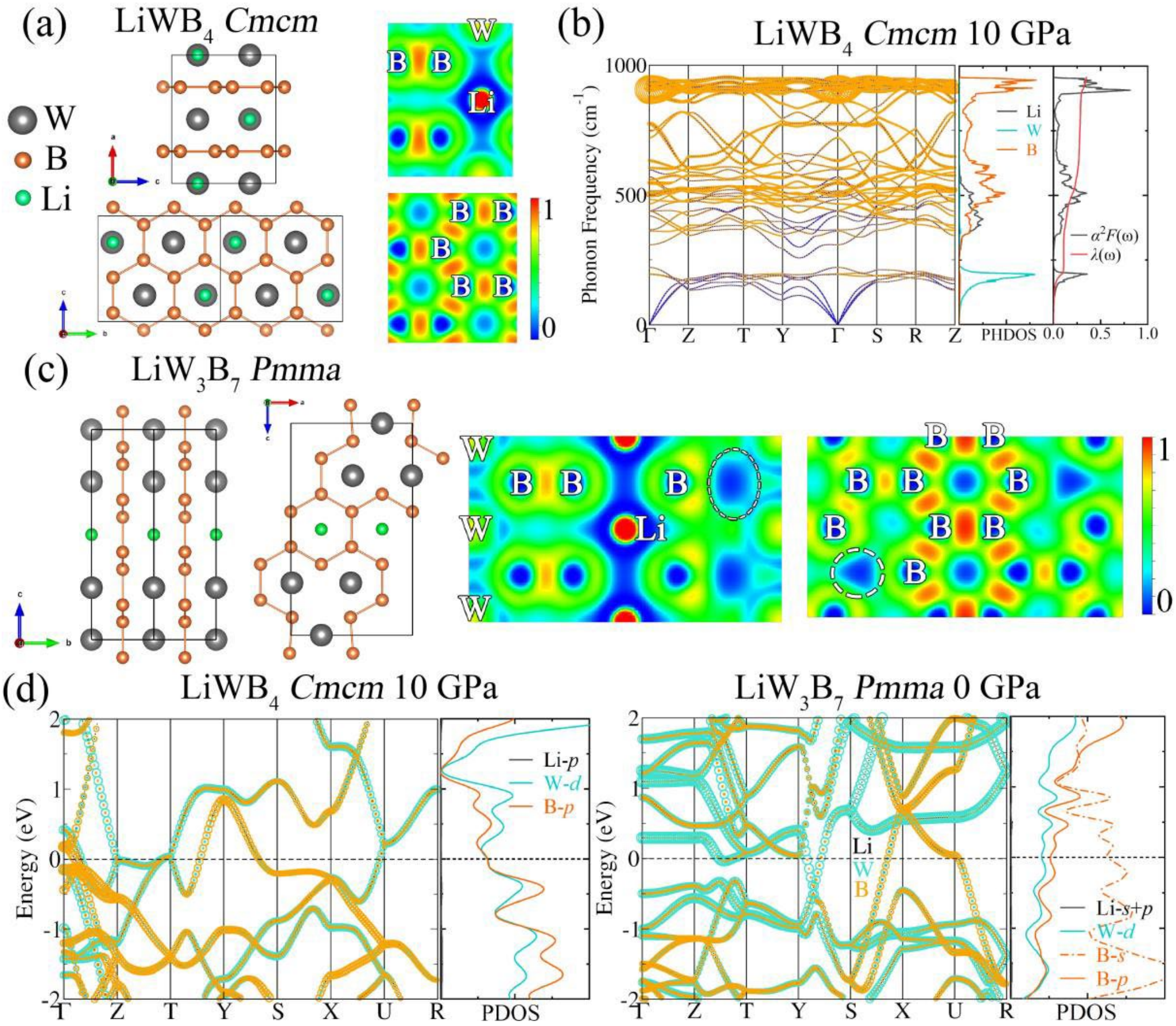

FIG. 5 (a) The crystal structure of the predicted phase $LiWB_4$ *Cmcm* and ELF at the (010) plane and the boron layer. (b) The calculated phonon curves, PHDOS, Eliashberg spectral function $\alpha^2F(\omega)$, the electron phonon integral $\lambda(\omega)$ of $LiWB_4$ *Cmcm* at 10 GPa. The orange solid dots represents the EPC constant $\lambda$ and the radii are proportional to the strength. (c) The crystal structure and ELF at the (-100) plane and the boron layer of the predicted phase $LiW_3B_7$ *Pmma*. (d) The band structures and partial density of states (PDOS) of $LiWB_4$ *Cmcm* and $LiW_3B_7$ *Pmma*.

The predicted $LiW_3B_7$ *Pmma* is dynamically stable ranging from 0 GPa to 60 GPa [Fig. S3] [70], which is also a layered structure composing of B layers and Li-W layers [Fig. 4(c)]. Notably, we can observe some cracks in the B layers, which breaks the six-fold rotation symmetry of the honeycomb B layer. We can also find these none-bonding features in the ELF, as marked in Fig. 4(c). In terms of the electronic structures [Fig. 5(d)], there is a saddle point around the *U* point along the Brillouin path, which is dominated by electrons of B atoms. Besides, the B-*s* electrons contribute significantly around the Fermi energy, suggesting distinct character from the role of B-*p* electrons in other predicted ternary compounds [Fig. 4 and Fig. 5(d)].

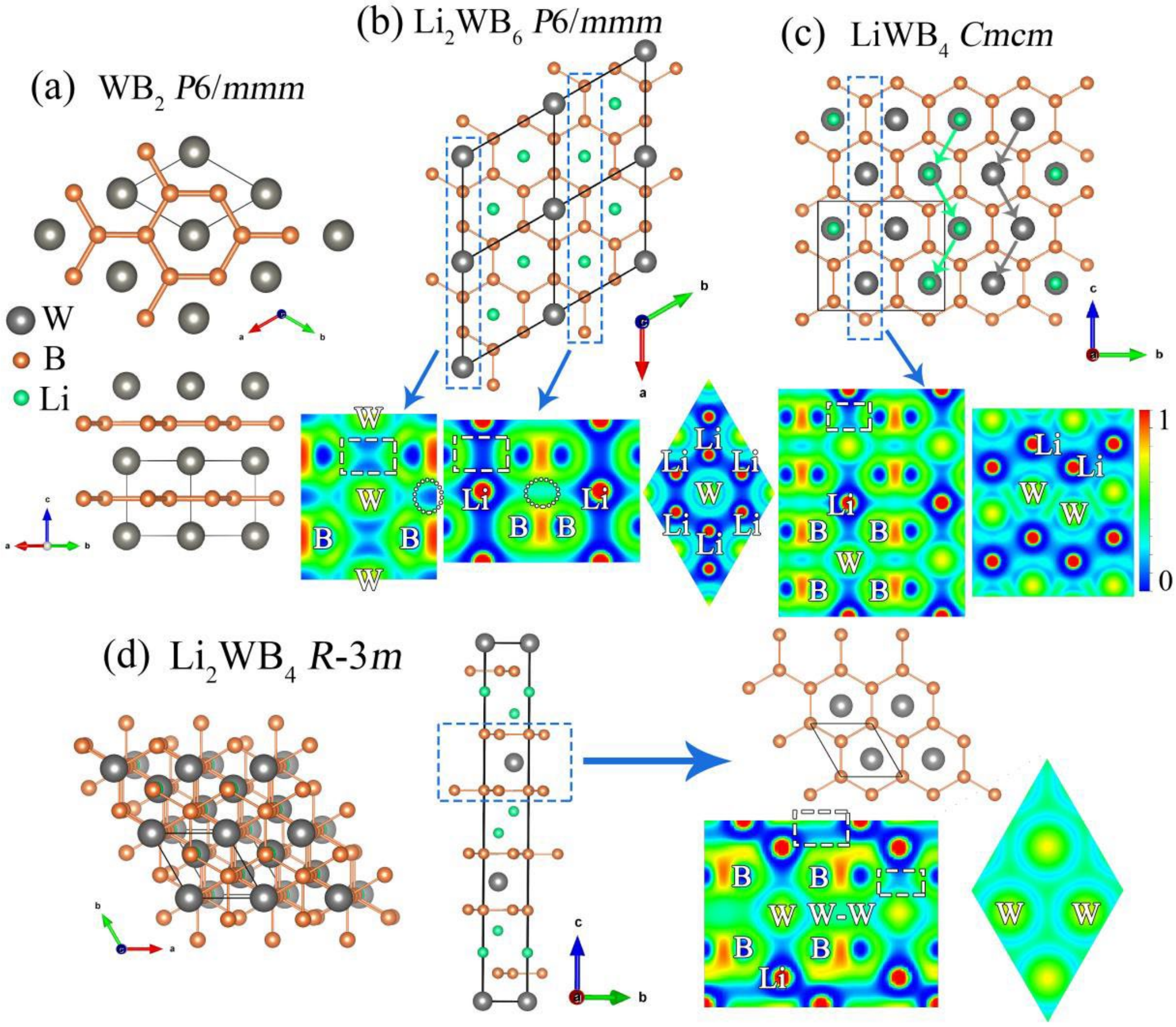

FIG. 6 (a) The crystal structure of $WB_2$ *P*6/*mmm*. (b) The crystal structure and ELF of $Li_2WB_6$ *P*6/*mmm*. (c) The crystal structures and ELF of $LiWB_4$ *Cmcm*. (d) The crystal structure of $Li_2WB_4$ *R*-3*m*, the $WB_4$ layer and the ELF.

Moreover, we perform the qualitative analysis on the superconductivity of the predicted ternary Li-W-B compounds from the perspective of bonding characteristics. According to Ref. 34 and Ref. 81, there are three crucial factors for the $T_c$ value of $WB_2$, including the proportion of flat honeycomb B layers, the coupling between W-*d* electrons and the W-W $\sigma$ bond stretching induced out-of-plane phonon modes. This could partly explain the absence of superconductivity in $Li_2WB_6$ *C*2/*m* and $LiW_3B_7$ *Pmma*, in which the buckling and the breaking in B layers invalidate the couplings that are conductive to superconductivity. Comparing the crystal structures of $WB_2$ *P*6/*mmm*, $Li_2WB_6$ *P*6/*mmm* and $LiWB_4$ *Cmcm* [Fig. 6 (a)-(c)], we can regard $Li_2WB_6$ *P*6/*mmm* and $LiWB_4$ *Cmcm* as the structures where the W atoms are substituted by Li atoms in $WB_2$ *P*6/*mmm*. The Li atoms replace the neighboring W atoms in Li-W layer and keep the space group in $Li_2WB_6$ *P*6/*mmm*, while the Li and W atoms form corresponding zigzag chains in $LiWB_4$ *Cmcm*. From the ELF results shown in Fig. 6(b), we can observe the covalent bonds in the center of the B honeycomb when sandwiching the W atoms, which resembles the condition in $WB_2$ *P*6/*mmm* [81]. Despite the Li atoms interrupt these center covalent bonds, the inter-layer bonding between B layers becomes more strengthened [Fig. 6(b)], which agrees with the PDOS that B-*p* electrons are predominant around the Fermi energy [Fig. 4]. Meanwhile, the Li atoms refine the formation of W-W $\sigma$ bond in the Li-W layers. Thus, Li atoms act as the dilution of W atoms in $WB_2$ *P*6/*mmm*, which restricts the W-W $\sigma$ bond but maintains the inter-layer couplings between B and W layers. As for $LiWB_4$ *Cmcm* shown in Fig. 6(c), the interval arrangement of Li-W-Li-W along the a-axis and b-axis limits both the bonds in the center of the B honeycomb layer and W-W $\sigma$ bond within the Li-W layer, which brings down the strength of both inter-layer and intra-layer couplings and results in the weak superconductivity of $LiWB_4$ *Cmcm* [Table. 2]. In Fig. 6(d), we picked out the B-W-B layer of $Li_2WB_4$ *R*-3*m*, which has the stoichiometry of $WB_4$. The enlarged ELF suggests the slight distribution between B layers and the center of the B honeycomb layer. Nevertheless, the ELF indicates covalent bonding in the W layers, suggesting clues of the W-W $\sigma$ bond. This is also in line with the main role of W-*d* electrons in PDOS [Fig. 4(c)]. Hence we believe that the Li atoms in $Li_2WB_4$ *R*-3*m* constrain the inter-layer coupling, while the $WB_4$ layer and the W-W bond retain the intra-layer coupling as $WB_2$ *P*6/*mmm*.

To further understand the superconducting properties of $Li_2WB_6$ *P*6/*mmm* and $Li_2WB_4$ *R*-3*m*, we compared the distances of B-B bonds and B-W-B layers in $Li_2WB_6$ *P*6/*mmm*, $Li_2WB_4$ *R*-3*m* and $WB_2$ *P*6/*mmm* under high pressure, as plotted in Fig. 7. $Li_2WB_6$ *P*6/*mmm* contains two types of B-B bonds, one is the (B-B)1 bond whose first neighboring atoms are two Li atoms surrounding the bond center, the other is the (B-B)2 bond whose bond center is surrounded by Li and W atoms [Fig. 7(a)]. In the aspect of B-B bonds [Fig. 7(b)], the (B-B)2 bond in $Li_2WB_6$ *P*6/*mmm* and B-B bond in $Li_2WB_4$ *R*-3*m* are comparable to that of $WB_2$ *P*6/*mmm*, while (B-B)1 bond in $Li_2WB_6$ *P*6/*mmm* is relatively pressurized. In particular, the layer distance of $Li_2WB_6$ *P*6/*mmm* [Fig. 7(c)] suggests that the inter-layer direction is more compressed, and the B-W-B distance at 0 GPa is comparable to that of $WB_2$ *P*6/*mmm* around 70 GPa. By contrast, the B-W-B distance indicates that $Li_2WB_4$ *R*-3*m* is more similar to $WB_2$ *P*6/*mmm* in Fig. 7(c).

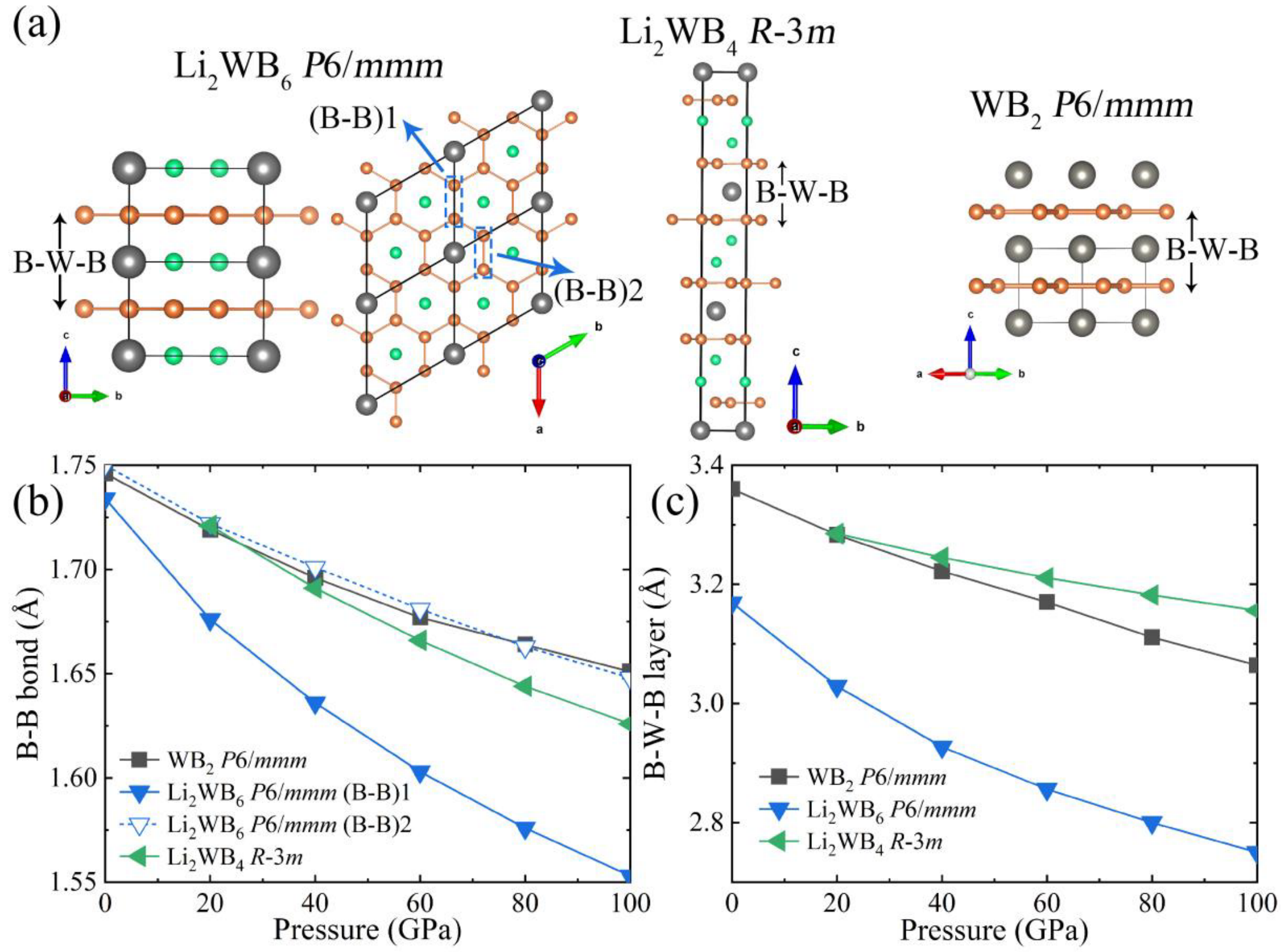

FIG. 7 (a) The B-W-B layers of $Li_2WB_6$ *P*6/*mmm*, $Li_2WB_4$ *R*-3*m* and $WB_2$ *P*6/*mmm*, and the bonds of (B-B)1 and (B-B)2 in $Li_2WB_6$ *P*6/*mmm*. (b) The distances of the B-B bond and the B-W-B layer of $Li_2WB_6$ *P*6/*mmm*, $Li_2WB_4$ *R*-3*m* and $WB_2$ *P*6/*mmm* under high pressure.

Therefore, combining the analysis above, the introduction of Li atoms induces the chemical pressure and enhance the inter-layer couplings in $Li_2WB_6$ *P*6/*mmm*, while Li atoms conserve the intra-layer couplings in $Li_2WB_4$ *R*-3*m*. On the one hand, both of the predicted structures lower the measured pressure of ~100 GPa in$WB_2$ *P*6/*mmm* [33, 34] to 0 GPa and 20 GPa, respectively, suggesting the potential for synthesizing and future experimental studies in Li-W-B system. On the other hand, our predicted structures illustrate that the flat B layers and role of metal elements in EPC are key factors in designing unique ternary borides under high pressure, and we expect more studies in ternary borides such as the introduction of alkali metals or transition metals to Ca-B system as well as Mg-B system via the strategy of TCLs, which may lower the stable pressure and explore more superconducting structures.

To evaluate the validity of the strategy of TCLs, we performed preliminary FTS in Li-W-B system at 0 GPa. As plotted in Fig. S5 (a) [70], we searched $Li_xW_yB_z$ [a/b ≤ 8 (a,b=x,y,z) and x+y+z ≤30] ternary compounds at 0 GPa, and the FTS almost covers regions within and out of TCLs. After comparing the energy with simple substances, the compositions with relative enthalpy ΔE < 0 from FTS and the results from TCLs are plotted in Fig. S5 (b) [70]. On the one hand, most of the compositions with $\Delta E < 0$ are around the low energy region, which are close to the results of TCLs. This suggests the validation of the second assumption of TCLs that current ternary convex hull reflects searching regions for potential stable and meta-stable compounds. On the other hand, although the FTS reproduce the meta-stable composition of $Li_2WB_6$ and $LiW_3B_7$ in TCLs, it did not find other thermodynamically stable ternary compositions, which has no influence on the meta-stability of the predicted Li-W-B compounds by TCLs.

Moreover, we can briefly estimate the computational cost through the calculated ternary compositions. Considering the formulas, there are overall 98 compositions calculated in the TCLs of Li-W-B at 0 GPa, while there are over 3000 compositions in the corresponding FTS, and we assume that the application of the TCLs strategy could improve the calculation efficiency by one or two orders of magnitude comparing with that of FTS. Therefore, the test of FTS in Li-W-B at 0 GPa provides evidence for the validation and efficiency of the TCLs strategy, and we expect to test more systems to improve the details and quality of this strategy in future work.

**B. Predictions of Li-Mo-B system.**

In the Li-Mo-B system, we carried out the structure searches along four TCLs including Li-MoB, Li-$MoB_2$, LiB-MoB and LiB-$MoB_2$, and we constructed the ternary convex hulls ranging from 0 GPa to 60 GPa [Fig. 8 (a)-(d)], combining the calculation results of enthalpy and phonon spectra [Fig. S4] [70]. The detailed enthalpy differences of the predicted compositions under high pressure are listed in Table. 3. The simple substances are the references of the ternary convex hull [66-69, 82]. We conducted structure searches in Li-W compounds but none of the predicted compounds have negative formation energy values compared with Li plus W.

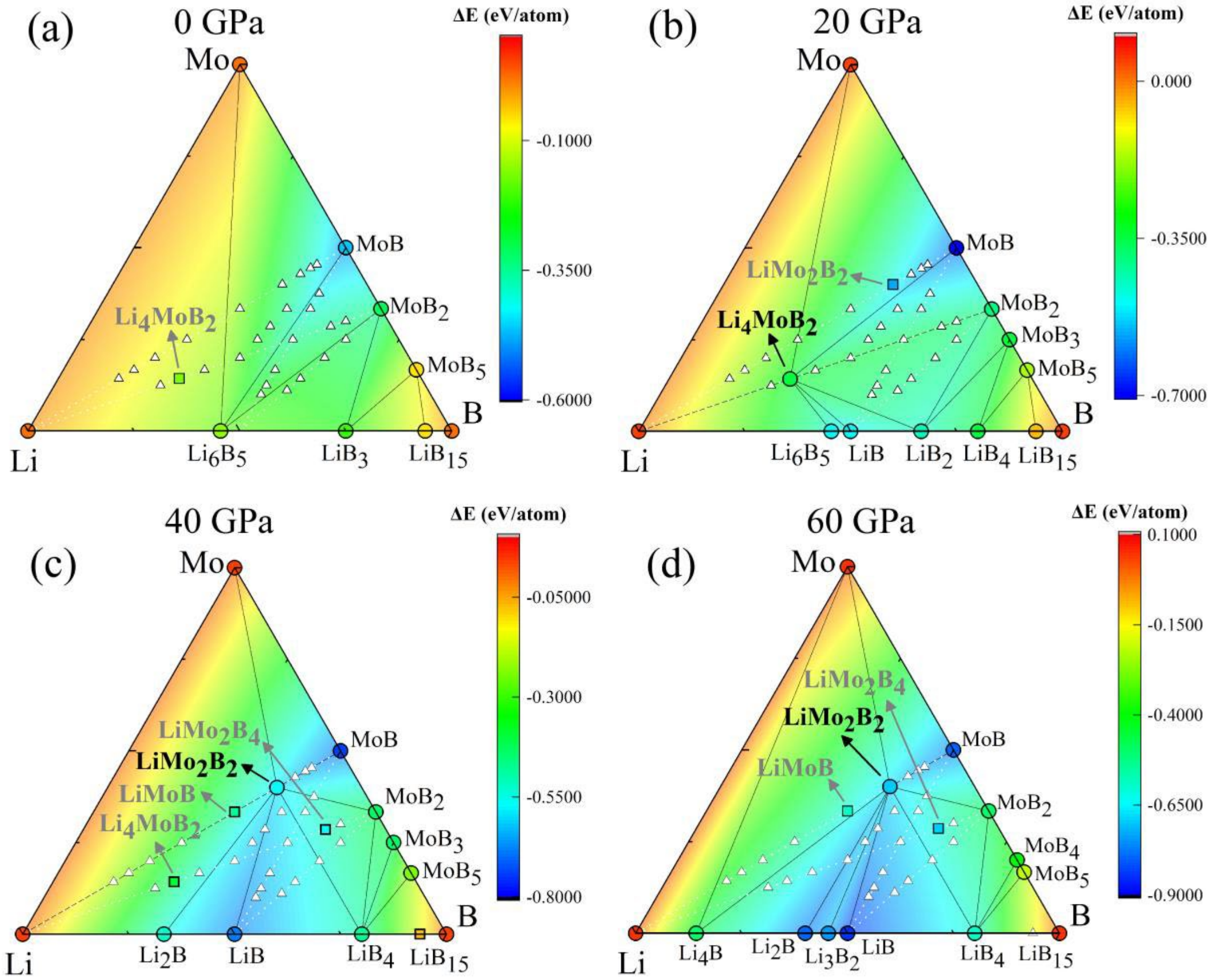

FIG. 8 (a)-(d) The ternary convex hulls of Li-Mo-B system under high pressures. The circle and black font are thermodynamically stable compositions, and the square and grep font are thermodynamically meta-stable compositions.

The composition of $Li_4MoB_2$ is thermodynamically stable around 20 GPa but exceeds the threshold of meta-stability (0.05 eV/atom) [59] around 60 GPa [Table. 3], which is mainly due to the modification of binary convex hull in Li-B [Fig. 8 (a)-(d)]. With the increasing pressure, the composition of $LiMo_2B_2$ becomes meta-stable around 20 GPa and is stable after 40 GPa [Fig. 8 (b)-(d)]. Although both $LiMo_2B_4$ and LiMoB are meta-stable above 40 GPa [Fig. 8 (c) and (d)], their dynamical stability differs from each other. $LiMo_2B_4$ hosts dynamic stability from 0 GPa, while LiMoB is dynamically unstable below 40 GPa [Fig. S4] [70].

TABLE 3. The enthalpy difference of the predicted Li-Mo-B compounds relative to the ternary convex hull under high pressure.

| ΔEnthalpy (eV/atom) | | | | | |
|---|---|---|---|---|---|
| Phase | Space Group | Pressure (GPa) | | | |
| | | 0 | 20 | 40 | 60 |
| $Li_4MoB_2$ | *C*2/*m* | 0.031 | 0 | 0.044 | 0.053 |
| $LiMo_2B_2$ | *Immm* | 0.164 | 0.027 | 0 | 0 |

| $LiMo_2B_4$ | *R*-3*m* | 0.138 | 0.083 | 0.024 | 0.008 |
|---|---|---|---|---|---|
| LiMoB | *Amm*2 | - | - | 0.021 | 0.041 |

Fig. 9 (a)-(d) illustrate the crystal structures of the predicted $Li_4MoB_2$ *C*2/*m*, $LiMo_2B_2$ *Immm*, $LiMo_2B_4$ *R*-3*m* and LiMoB *Amm*2. The $Li_4MoB_2$ *C*2/*m* is a monoclinic structure characterized by B zigzag chains along the b-axis [Fig. 9(a)]. $LiMo_2B_2$ *Immm* contains $B_2$ dimer surrounded by Li and Mo atoms [Fig. 9(b)]. $LiMo_2B_4$ *R*-3*m* is stacked by atom layers with Li atoms embedded in the center of the honeycomb B layers, which leads to the buckling in the B layers [Fig. 9(c)]. The structure of the orthorhombic LiMoB *Amm*2 approximates $Li_4MoB_2$ *C*2/*m* and also contains the B zigzag chains along the *a*-axis [Fig. 9(d)]. The detailed lattice parameters of the predicted Li-Mo-B structures are presented in Table S2 [70].

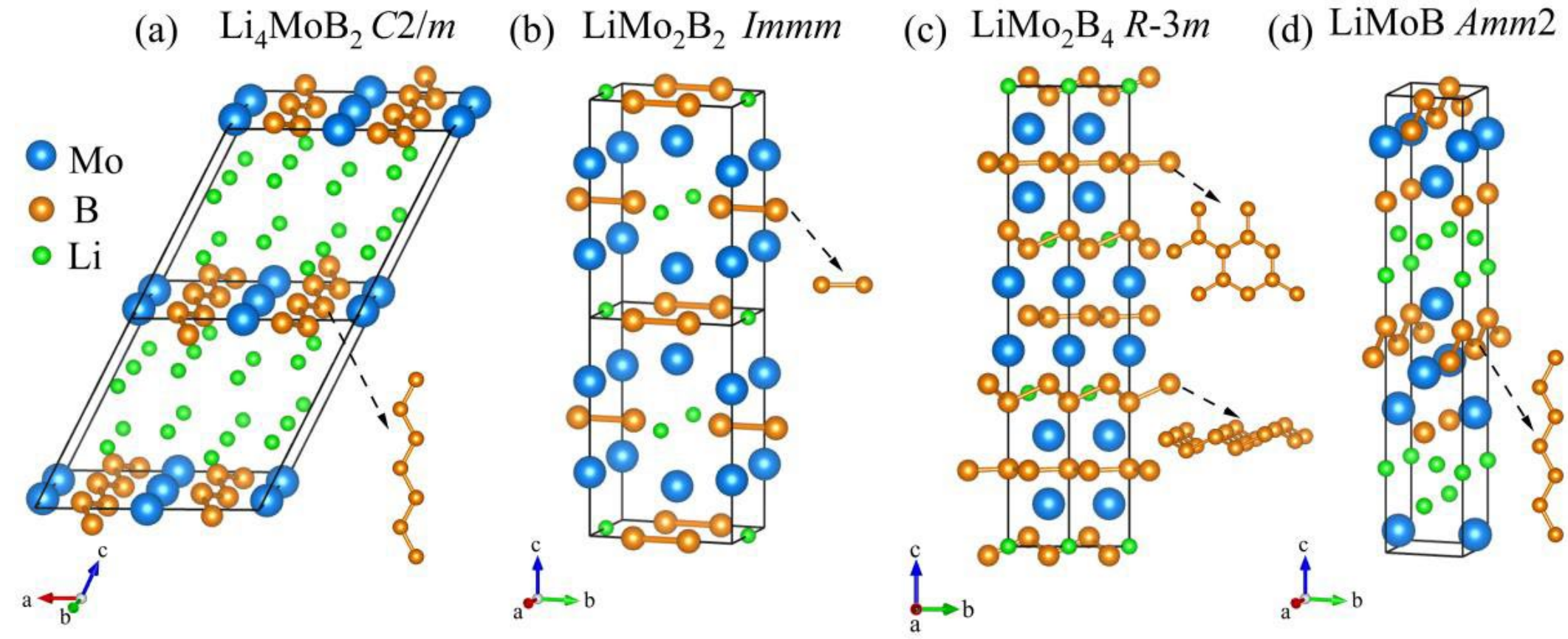

FIG. 9 The crystal structures of the predicted (a) $Li_4MoB_2$ *C*2/*m*, (b) $LiMo_2B_2$ *Immm*, (c) $LiMo_2B_4$ *R*-3*m* and (d) LiMoB *Amm*2.

Then we calculated the band structures and partial density of states (PDOS) of the four predicted Li-Mo-B compounds under high pressure, as illustrated in Fig. 10. We can observe typical bands crossing the Fermi energy, suggesting the metallicity of the predicted structures. Besides, the PDOS indicate that the Mo-*d* electrons play a significant role around the Fermi energy in all the predicted structures [Fig. 10 (a)-(d)], which is similar to the PDOS distribution in $\alpha$-$MoB_2$ and $\beta$-$MoB_2$ under high pressure [83]. In addition, the calculated ELF results are in Fig. S7 [70], we selected (001) and (010) planes for $Li_4MoB_2$ *C*2/*m*, $LiMo_2B_2$ *Immm* and $LiMo_2B_4$ *R*-3*m*, while (001) and (100) planes for LiMoB *Amm*2. We can find strong covalent bonding between B atoms and the ionic bonds around Li atoms. The ELF reflect the crystal structure characters like zigzag chains, dimers and honeycomb by B atoms.

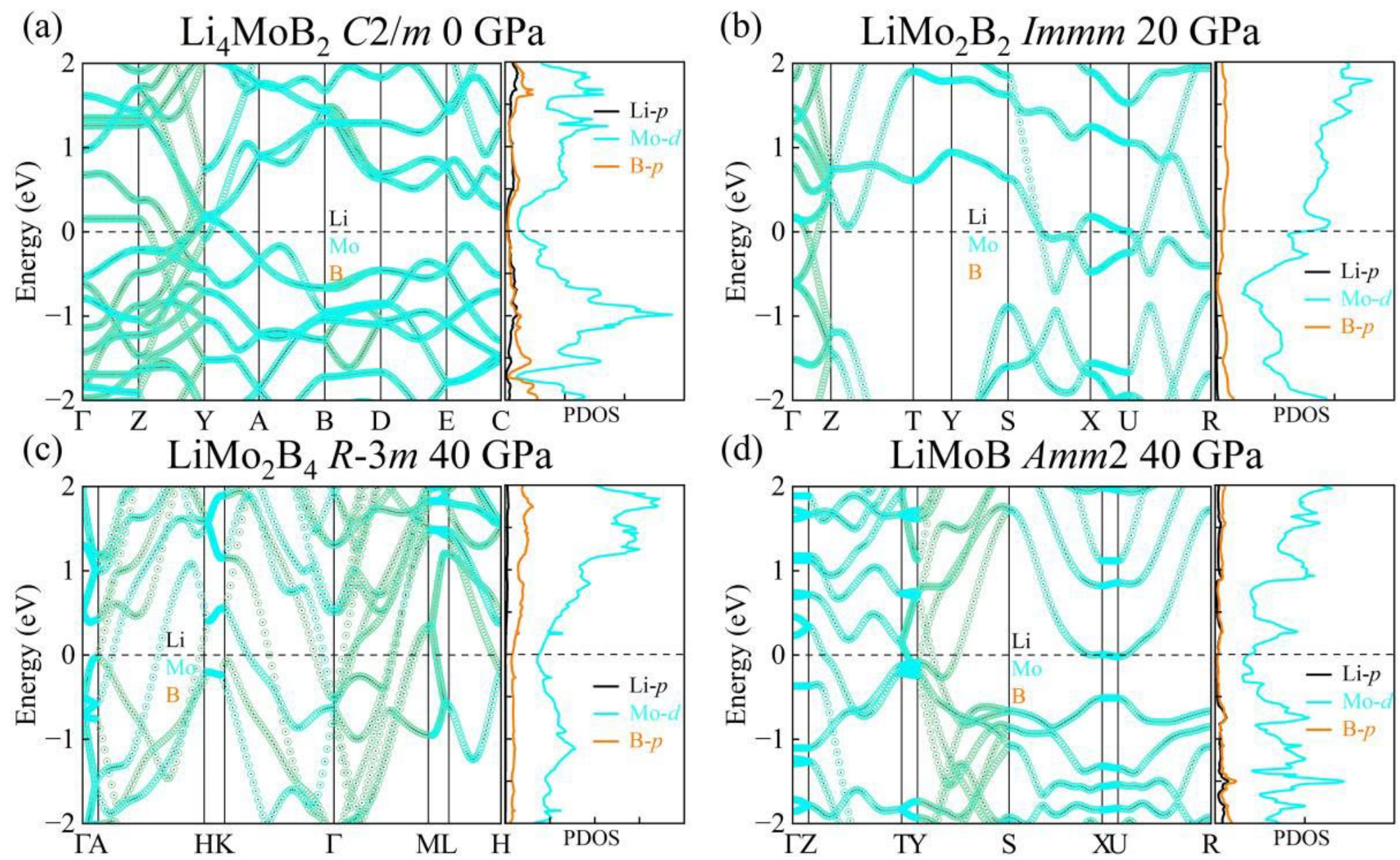

FIG. 10 The electronic structures and partial density of states (PDOS) of the predicted structures (a) $Li_4MoB_2$ *C*2/*m*, (b) $LiMo_2B_2$ *Immm*, (c) $LiMo_2B_4$ *R*-3*m* and (d) LiMoB *Amm*2 under high pressure.

Next we calculated the EPC properties for the predicted Li-Mo-B structures, among which the LiMoB *Amm*2 is not superconducting. As depicted in Fig. 11 (a)-(c), we can find that the Mo atoms associated vibrations modes are below 300 $cm^{-1}$, and they contribute significantly to the EPC in $Li_4MoB_2$ *C*2/*m*, $LiMo_2B_2$ *Immm* and $LiMo_2B_4$ *R*-3*m*. Accordingly, the Mo atoms account for about 66%, 73% and 69% to the integral $\lambda(\omega)$ in $Li_4MoB_2$ *C*2/*m*, $LiMo_2B_2$ *Immm* and $LiMo_2B_4$ *R*-3*m*, respectively. Given that the Mo-*d* electrons are dominant in the electronic states around the Fermi energy [Fig. 10], both the vibrations and electrons of Mo atoms are crucial for the EPC of the three predicted Li-Mo-B compounds. Nevertheless, the crystal structures of $Li_4MoB_2$ *C*2/*m*, $LiMo_2B_2$ *Immm* and $LiMo_2B_4$ *R*-3*m* mismatch the advantages for superconductivity, such as the flat B layers, the inter-layer or intra-layer couplings through the *d* electrons evidenced in the Li-W-B system. Hence all of the three structures demonstrate relatively weak superconductivity and the $\lambda$ is 0.25 with the $T_c$ value 0.2 K for $Li_4MoB_2$ *C*2/*m* at 0 GPa, $\lambda$ is 0.39 with the $T_c$ value 1.5 K for $LiMo_2B_2$ *Immm* at 20 GPa and $\lambda$ is 0.39 with the $T_c$ value 1.6 K for $LiMo_2B_4$ *R*-3*m* at 40 GPa, as listed in Table. 4.

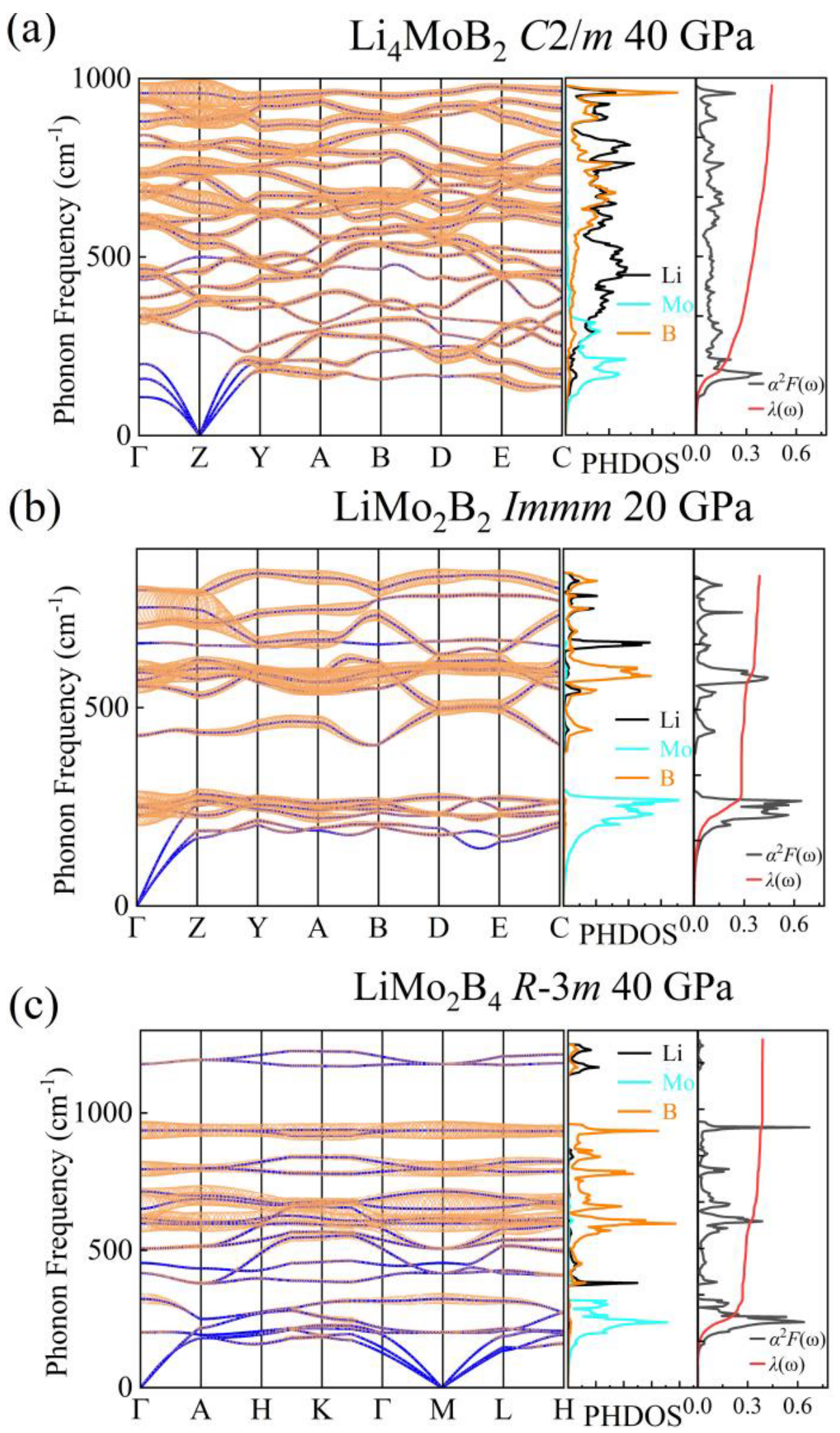

FIG. 11 The calculated phonon curves, PHDOS, Eliashberg spectral function $\alpha^2F(\omega)$, the electron phonon integral $\lambda(\omega)$ of (a) $Li_4MoB_2$ $C2/m$ at 40 GPa, (b) $LiMo_2B_2$ *Immm* at 20 GPa and (c) $LiMo_2B_4$ $R$-3$m$ at 40 GPa. The orange solid dots represents the EPC constant $\lambda$ and the radii are proportional to the strength.

In addition, we find that an anomalous enhancement of superconductivity under higher pressure in $Li_4MoB_2$ $C2/m$, whose $T_c$ value rises from 0.17 K at 0 GPa to 3.3 K at 40 GPa [Table. 4]. To understand such an order-of-magnitude increase, we calculated the electronic structures of $Li_4MoB_2$ $C2/m$ at 40 GPa [Fig. S8] [70]. The PDOS of Li and B atoms seem to be more notable than that at 0 GPa [Fig. 10(a)] and more electronic bands cross the Fermi energy such as along the Brillouin paths of $\Gamma$-$Z$, $B$-$D$ and $E$-$C$. Thus we compared the PDOS at Fermi energy ($N_{\mathrm{Ef}}$) for Li, Mo and B atoms under different pressures, as shown in Fig. S9(a) [70], and we can observe a sharp increase in $N_{\mathrm{Ef}}$ at 30 GPa, with the magnitudes increase about 41%, 82% and 77% than that at 20 GPa for Mo, Li and B atoms, respectively. This suggests that the electronic states from Li and B atoms are more notable under high pressure, which correlates with the rising of $T_c$. Besides, the results of Bader charge analysis [Table. S3] [70] provide further supports. The electrons donating in Li atoms and accepting in B atoms becomes more pronounced above 20 GPa, which agrees with the PDOS tendency under high pressure. Furthermore, we calculated the lattice parameters of $Li_4MoB_2$ $C2/m$ under high pressure [Fig. S9(b)] [70]. The c-axis direction is more pressurized than the other two directions, illustrating that anisotropic compression induces the enhancement of $T_c$ values via the variation of electronic structures.

TABLE.4 The calculated superconducting properties of the predicted $Li_4MoB_2$ $C2/m$, $LiMo_2B_2$ *Immm* and $LiMo_2B_4$ $R$-3$m$ under high pressure.

| | Space group | Pressure (GPa) | $\omega_{log}$ | $N_{Ef}$ | $\lambda$ | $T_c$ (K) $\mu^*$=0.1 |
|---|---|---|---|---|---|---|
| $Li_4MoB_2$ | $C2/m$ | 0 | 378.16 | 11.51 | 0.25 | 0.17 |
| | | 20 | 328.63 | 8.51 | 0.33 | 0.38 |
| | | 40 | 421.80 | 11.00 | 0.45 | 3.29 |
| $LiMo_2B_2$ | *Immm* | 20 | 425.06 | 12.05 | 0.39 | 1.51 |
| $LiMo_2B_4$ | $R$-3$m$ | 40 | 450.35 | 9.79 | 0.39 | 1.59 |

By investigating the Li-Mo-B ternary compounds through the strategy of TCLs, we promote the understanding of the Li-Mo-B phase diagram, preliminary indicating the reliability of TCLs strategy. Besides, we speculate that more 1$^{st}$ generation of TCLs and 2$^{nd}$ generation of TCLs in Li-Mo-B system may further explore superconducting ternary compounds such as along Li-$MoB_3$, Li-$MoB_5$, $LiMoB_4$-B, etc.

As for the synthesizing and superconductivity of the predicted Li-X-B (X=Mo, W) ternary compounds, we focused on the reported borides with both theoretical calculations and experimental measurements [31-34, 81, 84, 85]. The Kagome calcium boride $CaB_3$ is thermodynamically unstable with enthalpy over 0.2 eV/atom above the convex hull, which is synthesized by laser heating in diamond anvil cell (DAC) around 50 GPa [84]. The Mg-Fe-B ternary compounds are synthesized by mixing powder and heating in the ampoules, whose powder X-ray diffraction signal originate from the meta-stable phases of $Mg_2Fe_7B_7$ and $MgFe_3B_2$ [85]. $Mg_2Fe_7B_7$ and $MgFe_3B_2$ are 0.010 eV/atom and 0.055 eV/atom above the convex hull, respectively. In comparison, the predicted Li-X-B (X=Mo, W) ternary compounds have better thermodynamical stability than $CaB_3$ and are analogous to Mg-Fe-B compounds. In addition, meta-stable phases $Mg_2Fe_7B_7$ and $MgFe_3B_2$ are within the low energy region and are along the TCLs of Mg-FeB and $MgB_2$-Fe, which provides evidence for the prediction validity of the TCLs strategy. In particular, although $Li_2WB_6$ $P6/mmm$ is meta-stable in the Li-W-B ternary convex hull, it is stable within 60 GPa in the convex hull of LiB-$WB_4$, as shown in Fig. S6 [70]. This suggests that $Li_2WB_6$ *P*6/*mmm* could be synthesized by LiB+$WB_4$ under appropriate temperature and pressure condition.

In addition, we applied the same theoretical framework to calculate the superconducting properties of our predicted Li-X-B (X=Mo, W) ternary compounds as $\alpha$-$MoB_2$ [31, 32], $WB_2$ [33, 34, 81] and $CaB_3$ [84] under high pressure. The estimated $T_c$ values are ~33 K and ~17 K in $\alpha$-$MoB_2$ and $CaB_3$, respectively, which are overall in agreement with measured $T_c$ values ($T_c$ ~32 K for $\alpha$-$MoB_2$, $T_c$ ~22 K for $CaB_3$) [31, 32, 81]. Although there is difference between the calculated $T_c$ values (25-40 K) [81] and experimental values (~15-17 K) [33, 34], it comes from the meta-stable stacking faults and twin boundaries that resembles $MgB_2$ type structure under high pressure [33], which is key to the superconducting transition [81]. The predicted ternary compounds are the extension of Mo-B and W-B compounds, suggesting the expected $T_c$ values in the experiments.

Hence, we believe that our study in Li-X-B (X=Mo, W) ternary compounds is valid to guide the synthesizing and superconductivity measurements in the future experimental researches.

## IV. Conclusion

In summary, combining the first-principles calculations and the machine learning graph theory assisted universal structure searcher, we predicted ternary Li-X-B (X=Mo, W) compounds under high pressure via the strategy of TCLs. We predicted five unique structures in Li-W-B and four structures in Li-Mo-B. The compositions of $LiWB_4$, $Li_4MoB_2$ and $LiMo_2B_2$ could be stable in the ternary convex hull, while the other predicted compounds are meta-stable. Among these meta-stable structures, $Li_2WB_6$ *P*6/*mmm* and $Li_2WB_4$ *R*-3*m* are dynamically stable since 0 GPa and 20 GPa, respectively. Their superconductivity behaviour is analogous to $WB_6$ *P*6/*mmm* with $T_c$ values about 11 K, which greatly reduces the measured pressure from around 100 GPa in $WB_6$ *P*6/*mmm* to the mild pressure range. In $Li_2WB_6$ *P*6/*mmm*, the enthalpy difference relative to the convex hull at 0 GPa is around 0.02 eV/atom, suggesting high potential for experimental studies. We also find an abnormal increase of $T_c$ in $Li_4MoB_2$ *C*2/*m* under high pressure. Our studies shed light on the phase diagram of Li-X-B (X=Mo, W) under high pressure and provide theoretical evidence for the validity of the TCLs strategy in structure predictions. This strategy could be applied to the studies of unique ternary compounds with relatively limited computational resources, particular the structures that can further reduce the stable pressure and preserve the superconductivity. We anticipate refining the strategy of TCLs and integrating it with other techniques like artificial intelligence and high-throughput computing, thereby providing insights for the future experimental and theoretical works in unique ternary compounds.

## Acknowledgement

This work was supported by the National Natural Science Foundation of China (Grant Nos. 12474018, 52272265, 12204138, 12125404, T2495231, 123B2049, and 12504277), the National Key R&D Program of China (Grant No. 2022YFA1403201), the Advanced MaterialsNational Science and Technology Major Project (Grant No. 2024ZD0607000), the Basic Research Program of Jiangsu (Grant Nos. BK20233001 and BK20241253), the Jiangsu Funding Program for Excellent Postdoctoral Talent (Grant Nos. 2024ZB002, 2024ZB075, 2025ZB440 and 2025ZB852), the China Postdoctoral Science Foundation (Grant No. 2025M773331), the Postdoctoral Fellowship Program of CPSF (Grant No. GZC20240695 and GZC20252202), the AI & AI for Science Program of Nanjing University, Artificial Intelligence and Quantum physics (AIQ) program of Nanjing University, and the Fundamental Research Funds for the Central Universities. The calculations were supported by HPC platform of Shanghaitech University.

localization function results of predicted Li-Mo-B structures; the Bader charge results of predicted Li-Mo-B structures; the band structures of the predicted $Li_4MoB_2$ $C2/m$; the calculated partial density of states and lattice parameters of $Li_4MoB_2$ $C2/m$ under pressures. The Supplementary Material contains Refs. [71-74, 77-80].